# Low temperature photoluminescence study of polycrystalline ZnO

Sanjiv Kumar Tiwari[*]

Indian Institute of Technology Kanpur, India

We report on low temperature photoluminescence study of ZnO. The photoluminescence profile at 6K reveals $FX_A^{n=1}$, $FX_A^{n=2}$ and $FX_B^{n=1}$ peaks at 367.04 nm, 363.17 nm and 365.66 nm respectively together with neutral donor bound exciton $D^0{}_1X_A$ at 370.48 nm. The first, second, and third order longitudinal optical phonon replica of free exciton transitions $FX_A^{n=1}$ at 375.03, 383.10, and 391.54 nm respectively, and donor acceptor pair transition at 386.73 nm is observed. The dynamics of $D^0{}_1X_A$ and 1LO with excitation intensity is presented. A red shift of photoluminescence peak position of $D^0{}_1X_A$ and blue shift of 1LO with increase of pump intensity is observed. The peak photoluminescence intensity of $D^0{}_1X_A$ increases superlinearly with increase of pump intensity.

PACS numbers; 71.55.Gs, 78.30.Fs, 78.55.-m.

**Introduction**

Zinc oxide is a semi-conducting material used in variety of applications, ranging from sunscreen lotion and talcum powder to piezoelectric transducers and phosphors.[1,2] The band gap structure and optical properties of ZnO are quite similar to GaN, however, it has advantage over GaN in terms of strong binding energy of excitons e.g. 60 meV compared to 24 meV (GaN). High binding energy of exciton makes it potential candidates for optoelectronic devices and ultra violet LEDs as photoluminescence in ZnO has been reported at high temperatures.[3-5] The room temperature PL is dominated by the phonon replicas of the free exciton transition with the maxima at the 1LO phonon replica.[6] Raman spectra of ZnO shows red shift of 1LO peak attributed to local heating of sample with optical pulse,[7] we observed blue shift of 1LO peak in PL at 6K with increase of pump intensity. The present report focuses on dynamics of DBE peak and 1LO peak with increase of pump intensity at 6K. It is known that the conduction band of ZnO is predominantly *s* like and the valence band *p* like. Because of the lower symmetry of wurtzite ZnO, we get three twofold degenerate valence bands due to crystal field splitting and spin-orbit coupling. These bands are conventionally denoted as A ($\Gamma_7$), B ($\Gamma_9$) and C ($\Gamma_7$).[8] However, the symmetry of band structure of ZnO has remained controversial.[9,10] Valence band gap ordering of ZnO was observed by Reynolds et-al,[11] by measuring photoluminescence (PL) in presence of magnetic field. Their resolved spectra showed the additional fine structure of the excitons concluding valence band symmetry ordering of (A-$\Gamma_9$, B-$\Gamma_7$ and C-$\Gamma_7$) in ZnO similar to that observed in most other wurtzite II-VI structure and GaN. Dynamics of free excitons, bound excitons and LO

---

[*]Author to whom correspondence should be addressed; electronic mail: tiwarisanjiv@gmail.com

replicas with temperature has been reported by several authors.[4-7]

**Experimental setup**

ZnO sample were prepared by cold press of ZnO powder 99.99% (Koch-Light Laboratories Ltd- England) at pressure of 6 Ton and were sintered at 1000 $^0$C in air for five hours. In order to see the effect of sintering on the sample, we took X-ray diffraction (XRD) spectrum of the sample before and after sintering. No broadening is observed in various XRD profiles of the sintered samples; however a slight asymmetry is noticed. It is concluded although a small homogeneous strains may be present but inhomogeneous strains are absent in the sintered samples used. The average grain size of 19 nm and 28 nm, before and after sintering respectively is estimated. Steady-state photoluminescence was carried out with the sample placed in a closed cycle cryostat (ARS, Model 830. USA) in the temperature range of 6-250 K. Third harmonic of Nd: YAG (DCR-4G, Spectra Physics) λ=355 nm, 5 nano-second pulse duration (FWHM), 10 Hz repetition rate was used as excitation source. The photoluminescence (PL) light was imaged onto fiber coupled monochromator and detected by intensified charge coupled device (ICCD DH720, Andor Technology).

**Results and discussion**

Figure 1(a) shows the PL profile of ZnO at various temperatures. As the temperature increases the discrete peaks of $D^0_1X_A$ and longitudinal optical phonon, 1LO get broader and finally merge into a single broad peak. The observed discrete peaks of free exciton, donor bound exciton (DBE), donor acceptor pair (DAP) and LO replicas at 6K are shown in figure 1(b). Since the peaks are well resolved at 6K, the PL at 6 K was chosen to study the dynamics of DBE and 1LO replica with increase of pump intensity. The emissive process from $\Gamma_7$ of conduction band to $\Gamma_9$ of valence band is written as $\Gamma_7 \times \Gamma_9 = \Gamma_5 + \Gamma_6$, the $\Gamma_5$ exciton transition is allowed whereas the $\Gamma_6$ is forbidden. In Fig1 (b) the free excitonic transition is observed due to energy state n=1 and first excited energy state n=2 at 367.04 nm (3.378 eV) and 363.17 nm (3.414 eV) respectively. The value of free excitonic transition due to energy state n=1 at 367.04 nm (3.378 eV) is close to the predicted value of $FX^{n=1}_A$[12, 13]. So, the effect of homogeneous strain is negligible in our samples where the absence of inhomogeneous strain is already confirmed from the XRD analysis. Following the reported energy separation[14, 15] of A and B excitons, we have assigned the emission peak at 365.66 nm (3.391 eV) due to B exciton ($FX^{n=1}_B$), in agreement with the predicted value of longitudinal B exciton energy.[15] The measured energy separation of A exciton ($FX^{n=1}_A$) and B exciton is 13 meV, close to predicted experimental value of 9-15 meV.[16] The absence of the spectral signature of the C exciton associated with the $\Gamma_7$ interband transition is due to thermalisation down to lower exciton levels at low and room temperatures. The most intense PL peak at 370.48 nm (3.347 eV) is due to neutral donor bound exciton, $D^0_1X_A$. The other peaks observed on higher wavelength side of main PL profile are assigned as LO replicas of $FX^{n=1}_A$. These phonon replicas are very weak and mixed with other closely spaced peaks. We were able to resolve only 1LO phonon replica of free exciton which appears at 375.03 nm (3.306 eV) at 6K. The experimental detection of second order LO replicas and DAP is difficult because of their energy position in the spectral region where the donor acceptor transition (DAP) and 2LO replica of free exciton occurs. The two small shoulders appearing at 383.10 nm (3.236 eV) and 386.73 nm (3.206 eV) were assigned as 2LO replica and DAP transition of free exciton

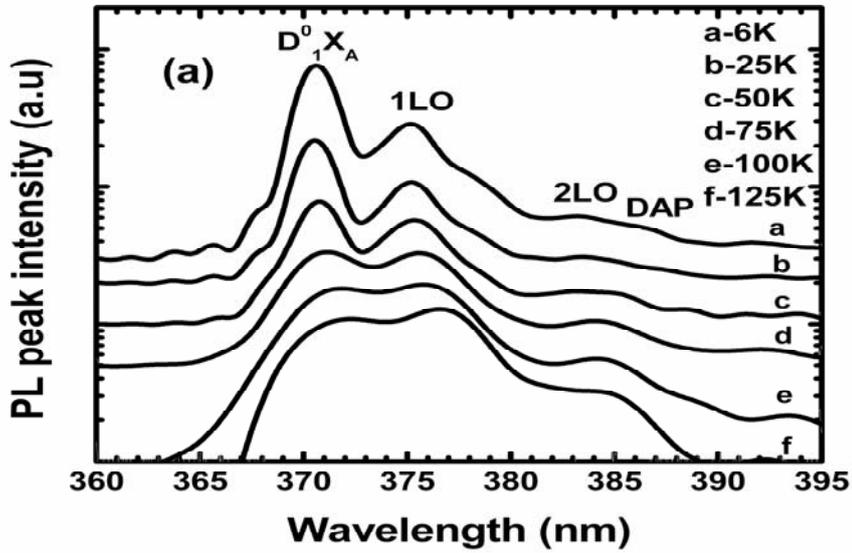

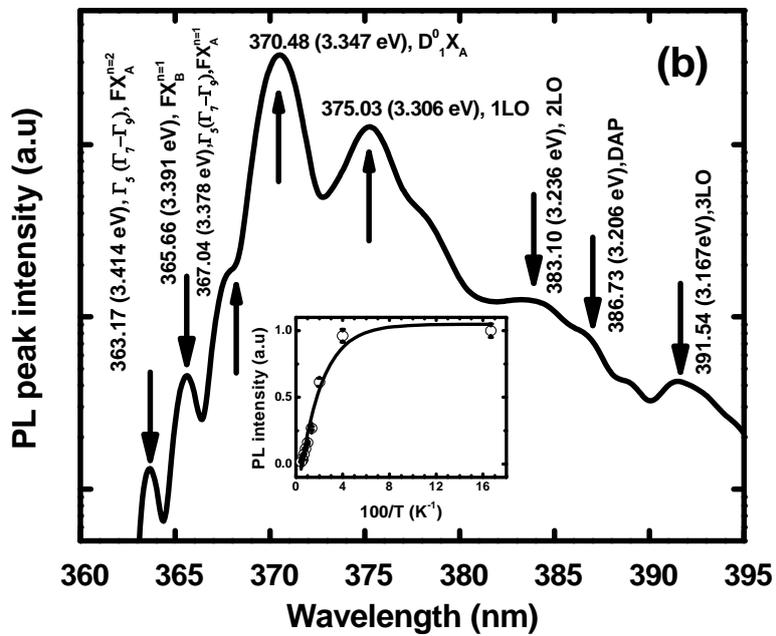

FIG.1. PL profile of bulk ZnO at excitation intensity 33 kW-cm$^{-2}$ (a) PL profile at various temperatures, and (b) PL profile at 6K. Inset of (b) shows the variation of Intensity with temperature

. To understand the behavior of DAP peak with temperature, consider the DAP peak position[17] given by,

$$\hbar\omega = E_g - \left\{ (E_D + E_A) - \frac{e^2}{\varepsilon r_{DA}} - \left(\frac{e^2}{\varepsilon}\right)\left(\frac{a}{r_{DA}}\right)^6 \right\}$$

where $\varepsilon$ is dielectric constant, $a$ the effective Vander Waals coefficient for the interaction between neutral donor and neutral acceptor, and $r_{DA}$ is distance between neutral donor and acceptor respectively. The rise in temperature due to increase in pump intensity will cause increase in $r_{DA}$ and hence as the temperature increases DAP peak moves away from 2LO peak and eventually disappears.

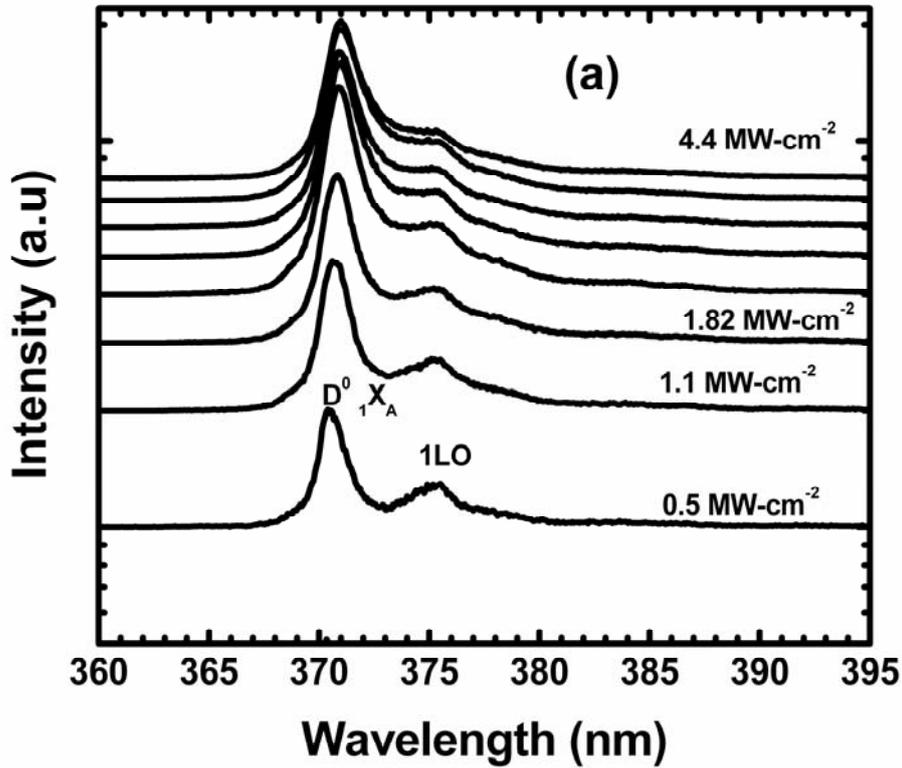

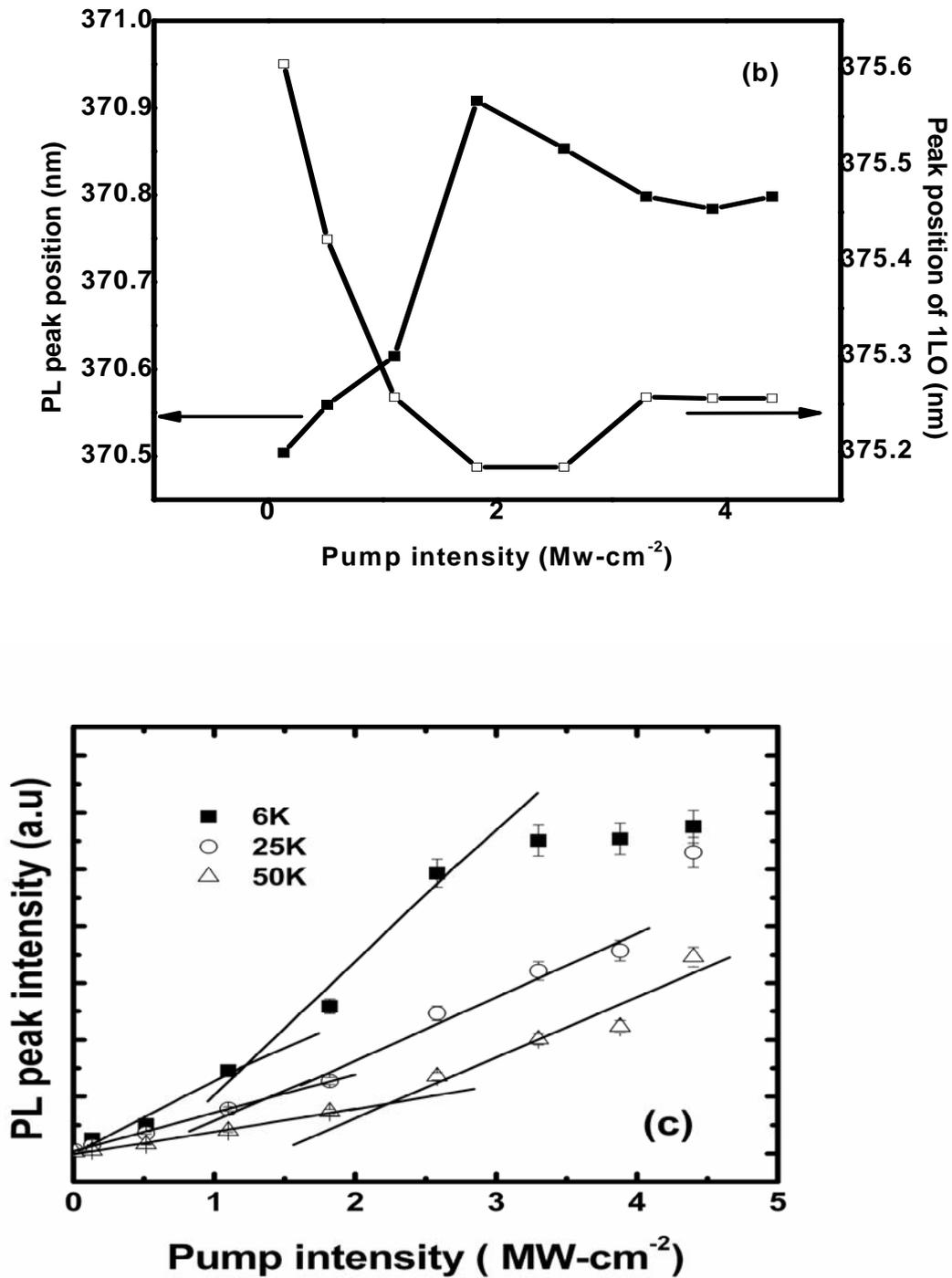

FIG. 2. Variation of PL peak intensity with increase of pump intensity at fixed temperature of 6K, (a) PL profile at various pump intensity, (b) Variation of PL peak position of DBE and 1LO with pump intensity, and (c) variation of PL peak intensity of DBE with pump intensity at 6, 25, and 50K

The PL peak positions are determined by averaging the PL profile over ten points. Each point on the curve is an average of ten observations. Figure 2 shows the variation of DBE peak $D^0{}_1X_A$ with increase of pump intensity. It is evident from figure 2(b) that PL peak position below 2 MW-cm$^{-2}$ of excitation intensity emission spectra is red shifted and 1LO of free excitons blue shifted. However, above the pump intensity of 2 MW-cm$^{-2}$ there is no significant variation of peak positions on further increase of pump intensity. This behavior may be due to heating effect of sample with pump intensity.[18] Figure 2(c) shows the variation of PL peak intensity with excitation intensity at various temperatures. The PL intensity increases linearly up to 1MW-cm$^{-2}$ of pump intensity; however, beyond 1MW-cm$^{-2}$ of pump intensity the PL peak intensity increases super linearly. It should be noted that at 6K the $D^0{}_1X_A$ peak predominates. It is known that thermal behavior and radiative recombination of $D^0{}_1X_A$ and of $FX^{n=1}_A$ are same provided thermal energy is less than binding energy of neutral donor.[19] The binding energy of excitons with neutral donor can be estimated [20] from the theoretical fit of equation $I(T) = \dfrac{A}{1 + C \exp(-E_a/k_B T)}$, where A and C are temperature independent constants, I(T) is PL peak intensity at temperature T, $k_B$, is Boltzmann constant and $E_a$ is transition energy/activation energy at which free exciton quenched into bound exciton. i.e binding energy of excitons with neutral donor. The value of ($E_a$) obtained from the theoretical fitting as shown in inset of figure 1(b) is 10 meV (~120 K). However, in figure 1(b) the spectral energy difference between $FX_A{}^{n=1}$ and $D^0{}_1X_A$ is 31 meV, this is due to heating of sample with laser pulse. The predicted value energy separation between $FX_A{}^{n=1}$ and $D^0{}_1X_A$ is 12-15 meV.[16]

In order to understand dynamics of DBE with pump intensity, we assume that the rate equation describing the emission intensity at steady state,[21] $I_{PL} = \dfrac{n_x}{\tau_{FX}} + c_1 n_x^2$. Here $n_x$ is the density of excitons, which is proportional to the excitation intensity, and $I_{PL}$ is the rate of generation of exciton. $n_x/\tau_{FX}$ is the radiative recombination rate of free excitons whose radiative life time is $\tau_{FX}$, and the second term $c_1 n_x^2$ is the radiative recombination rate due to exciton-exciton interaction process. The decrease in threshold value of pump intensity where super linear increase of PL sets in decreases with decrease in temperature is attributed to enhancement of absorption process at lower temperatures.

The blue shift of 1LO peak is due to increase of free exciton thermal energy by pump intensity. A general relationship for the emission lines involving phonon and exciton emission can be written as $E_n = E_0 - (n\hbar\omega_{LO} - \Delta E)$, where $\hbar\omega_{LO}$ =72 meV (fig-1(b)) is LO phonon energy, $E_0$ is exciton energy and $\Delta E = \dfrac{\hbar^2 k^2}{2M}$ is kinetic energy of free exciton and is equivalent to $\dfrac{1}{2}k_B T$.[6] However, the energy separation between the $FX_A{}^{n=1}$ and 1LO emission peak exhibit a strong temperature dependence. The spectral line shape I(LO) of the first LO replica line of the free exciton has the form $I(LO) \approx E^{1/2} \exp(-\dfrac{E}{k_B T}) P(E)$ here E equals $\hbar\omega - (E_{FX_A^{n=1}} - \hbar\omega_{LO})$ and P(E) is the transition probability which varies as $E^m$ for the corresponding phonon assisted transition, where m is constant. At low temperatures (< 75K) P(E) is typically assumed to be proportional to E.[22]

Therefore, the peak position of 1LO band will be shifted by $3/2k_BT$ to the higher energy side. In our case, figure 2(b), 1 LO peak gets blue shifted by 3 meV, equivalent to 6K.

**Conclusion**

In conclusion PL at 6 K is dominated by donor bound exciton peak. PL peak corresponding to DBE shifts to higher wavelength side and 1LO peak of free exciton gets blue shifted with increase in pump intensity. This behavior is explained in terms of increase of free exciton thermal energy.